# Silicon Photonics: The Inside Story

Bahram Jalali

University of California, Los Angeles

Department of Electrical Engineering

Los Angeles, CA 90095-1594

jalali@ucla.edu

**Abstract:**
The electronic chip industry embodies the height of technological sophistication and economics of scale. Fabricating inexpensive photonic components by leveraging this mighty manufacturing infrastructure has fueled intense interest in silicon photonics. If it can be done economically and in an energy efficient manner, empowering silicon with optical functionality will bring optical communications to the realm of computers where limitations of metallic interconnects are threatening the industry's future. The field is making stunning progress and stands to have a bright future, as long as the community recognizes the real challenges, and maintains an open mind with respect to its applications. This talk will review recent 'game changing' developments and discuss promising applications beyond data communication. It will conclude with recent observation of extreme-value statistical behavior in silicon photonics, a powerful example of how scientific discoveries can unexpectedly emerge in the course of technology development.

**Introduction:**
Persuading silicon to perform photonic functions can bring optical communication to the world of chip interconnects. Envisioned are silicon chip that communicates internally, or with other chips, using photons, to avoid the bandwidth limitations imposed by metallic interconnects [1]. Other opportunities abound, including low cost transceivers for 10 to 100Gbit Ethernet, a new platform for mid-infrared photonics, and optically assisted analog-to-digital conversion, just to name a few.

Guided by such visions and propelled by pioneering research conducted in the 1980s and 1990s, silicon photonics has enjoyed spectacular progress in the last eight years. The critical size of photonic devices has been scaled to the 200 nm regime, an order of magnitude less than a decade ago, and at the limit imposed by the optical wavelength. Optical amplification and lasing, once considered forbidden in silicon, have been achieved [2]. High speed and efficient electrical-to-optical and optical-to-electrical conversion is being performed by production-worthy silicon devices. Silicon's nonlinear optical properties, enhanced by tight optical confinement in $Si/SiO_2$ structures, are producing wavelength generation and conversion - central functions in multi-wavelength communications and signal processing. Much of these developments are documented in recently published review articles [3-4]. Not meant as a comprehensive review, the present commentary aims to elucidate the fundamental trends and challenges of this exciting field. It also provides a sampling of the most recent research with emphasize on those that address the fundamental problems.

**Fundamentals:**
The highest impact of silicon photonics is believed to be in data communication, consequently, most of the research has been aimed at producing a source of photons, an electro-optic modulator, and a photodetector. Optical amplifiers are also needed to compensate for path losses, and to take full advantage of the optical bandwidth, wavelength manipulation devices will be desired to implement multi wavelength data channels and optical packet switching. Such devices include passive wavelength filters that combine or separate wavelength channels and active devices for wavelength conversion and switching.

It has been known since the early 1990s that one can create high quality (low loss and single mode) optical waveguides on silicon-on-insulator (SOI) wafers [5]. It was pointed out even earlier by Soref that one can attenuate light, and shift its phase, by free carrier injection (or depletion) [6]. One can electrically modulate the photon intensity and hence, encode electrical data onto an optical carrier wave. With single mode waveguides, one can create interference based devices that combine or split light into different channels. During the past 25 years, a myriad of such devices have been demonstrated with recent activity judiciously aimed at reducing waveguide cross section from several microns to hundreds of nanometers.

Silicon is not able to detect signals at communication wavelengths of 1300nm and 1550nm because such photon energies are less than the bandgap. Yet these wavelengths represent standard communication bands because optical fiber, to which devices must eventually interface, has low propagation losses in these bands. Silicon's inability to absorb these wavelengths has been overcome by taking advantage of the small bandgap of germanium grown on silicon [7].

Coaxing silicon to emit and amplify light, and to change its wavelength, is far more difficult because of hard limits imposed by nature. Silicon is an inefficient light emitter because of its indirect bandgap and also because of free carrier absorption. The vast majority (but not all) of electron-hole pairs, created by electrical or optical injection, lose their energy to heat before they can create a photon. To make matters worse, injected free carriers will absorb photons through free carrier absorption. Any light that is created, or enters from outside, will not be amplified because the rate of stimulated emission is far below the rate of absorption by carriers [2,4]. Amplification is the pre-requisite to lasing, and that's why it was so difficult to make silicon lase using conventional approaches. A second fundamental problem with silicon is its lack of a static dipole moment, a consequence of its centrosymmetric crystal structure. This means that the linear electro-optic (Pockel) effect, that wonderful phenomenon that makes LiNbO3 (Lithium Niobate) and III-V semiconductors good electro-optic materials, is absent in silicon. Today, the only practical way to encode data onto light is by modulating its absorption via carrier (electrons and holes) injection or depletion.

This approach works well, particularly at low modulation speed where charge storage effects and junction capacitance can be ignored. It has been used since the late 1990's to create commercial optical attenuators that are used to adjust the power in multi wavelength fiber networks. Figure 1 is an example of a multichannel product from Kotura Corporation, a Southern California company. The product is deployed in metropolitan area networks by major carriers. High speed versions of these devices, aimed at electro-optic modulation for data transmission, have been reported by Luxtera Corporation, Intel, and others.

Much of the current research activity is focused on finding ways around silicon's physical limitations. Conclusive demonstration of optical amplification and the optically-pumped lasing were demonstrated in 2003-2004 by exploiting the Raman scattering, a phenomenon that describes interaction of light with atomic vibrations of the crystal [2]. Today, the question is no longer whether silicon can amplify light or lase, but rather how well. Unlike other nonlinear optical effects that can also lead to optical amplification, such as four wave mixing (FWM), phase matching is not needed in Raman devices. Phase matching is a convoluted term that describes the momentum conservation condition that must be satisfied before efficient transfer of energy from one wave to another can take place. It is difficult to achieve in practice; therefore, Raman devices are attractive because they don't require it. However, a limitation of Raman is the narrow gain bandwidth. Unless a broadband or multi-wavelength pump is used, the intrinsic 100 GHz gain spectrum of Raman is only sufficient for amplifying two 10Gbit/s data channels spaced by the industry standard 50GHz optical frequency spacing. Motivated by this, researchers have recently demonstrated broadband optical amplification by using FWM [8]. FWM occurs because of Kerr effect, a $3^{rd}$ order nonlinearity that exist in silicon, despite the material's lack of $2^{nd}$ order nonlinearity (Pockel effect). Because FWM requires phase matching, precise control over waveguide dimensions is necessary to achieve amplification using this approach, a requirement that will be difficult to achieve in production. To overcome such problems, electrical control over phase matching has recently been demonstrated [9]. This approach uses the elasto-optic effect realized with integrated piezoelectric transducers.

Raman and Kerr effects are examples of nonlinear optical phenomena which only appear when silicon is pumped with high intensity laser light $\geq$ (10 MW/cm$^2$). To avoid generating electrons and to prevent free carrier absorption, these devices use infrared light (1500 nanometer), so that the photon energy is less than the bandgap. However, because of high intensities involved, two-photon absorption (TPA) occurs (Fig. 2). Photons are lost not only because of TPA, but even more so through absorption by the generated free carriers. To avoid them, devices typically operate in the pulsed mode where electron accumulation does not occur. Because the pulse repetition period must be larger than the minority carrier lifetime (1-10ns in SOI waveguides), the utility of this technique is limited to low data rates. To operate continuously, one can apply an electric field (using a reverse biased p-n diode) to sweep out the electrons. However, this approach is only partially effective because the rate at which electrons can be swept out is limited by the saturated drift velocity ($10^7$ cm/s). Additionally, the diode results in electrical power dissipation on the chip. Hence, dealing with the TPA induced free carrier absorption remains the central focus of silicon photonics research. Recently, two photon photovoltaic effect was demonstrated, and was used to sweep out the TPA generated carrier while boasting negative electrical power dissipation [4]. The negative dissipation is achieved because the device harvests the energy of photons lost to TPA. In another potentially promising direction, helium bombardment has been used to reduce the accumulation of generated carriers by lowering the minority carrier lifetime [10]. This approach is inspired by lifetime reduction techniques used in silicon step recovery diodes: fast switching devices that are used for frequency comb generation.

The TPA is not always detrimental; it can, in fact, be exploited to perform useful functions. For example, a TPA photodetector can be used for monitoring the optical power in an integrated multi-wavelength channel equalizer [11]. It can also be used as a pulse compressor in an external cavity mode locked laser [12].

Significant excitement was created in 2000 by the report of internal optical gain (no net gain) from silicon nanocrystals embedded in SiO2 host. While obtaining net gain is frustrated by carrier absorption and scattering loss caused by nanocrystals, and the approach has not produced a laser yet, electroluminescent diodes that produce spontaneous emission continue to be improved [13]. Among recent ideas are the cavity enhancement of emission, and the use of silicon nitride as the host for nanocrystals (instead of SiO2) [14]. By lowering the tunneling barrier and hence, the operating voltage, the latter may enhance the device reliability and solve a lingering issue with such devices.

Japanese researchers at Hitachi have reported an electroluminescent device with a 9nm SOI active region [15] and a CMOS compatible structure. Dubbed a Light Emitting Transistor (LET), it can emit and detect light. Progress is also being made in hybrid silicon/III-V technologies with the preferred approach being the bonding of an indium-phosphide gain region onto a passive silicon microcavity. Using this approach, a joint Intel-UCSB team has been developing an electrically pumped hybrid laser [16].

**Future trends and Challenges:**
Future research must be guided by the fundamental challenges that must be surmounted before the vision of optically empowered CMOS technology is realized. For photonics to be truly compatible with VLSI chips, it must be compliant with the two basic requirements of the semiconductor industry: power efficiency and real estate efficiency. Power dissipation is the most pressing problem in the semiconductor industry, to the extent that it finally forced the microprocessor manufacturers to abandon higher clock speeds in favor of multi-core architectures [17]. Described by the Moore's Law, transistor count has been increasing at an astonishing pace with the today's microprocessors (Intel Itanium II), boasting nearly 2 billion transistors. The unintended consequence is that power density now exceeds 100 Watts/cm$^2$, an order of magnitude higher than a typical hot plate (Figure 3 [18]). This poses severe challenges for photonic devices, for they must be able to operate reliably at substrate temperatures as high as 90C. Diode lasers, such as those used in fiber optic communication, are packaged with a thermoelectric cooler because their performance severely degrades at high temperatures. It remains to be seen whether such type of lasers, regardless of what material they are made of (silicon, III-V, or hybrid), can efficiently and reliably operate on a VLSI chip. Far from being a compromise, today's use of an off-chip in silicon photonics is an intelligent solution. It does not subject the laser to the hot environment of a VLSI chip and also it does not add additional heat to the chip. Other photonic devices, such as the electro-optical modulators and optical amplifiers, reside on the chip and must be able to operate at elevated temperatures. High temperature compliance notwithstanding, photonic devices cannot dissipate large amounts power on their own, otherwise they exacerbate chip heating problem. Time has come for silicon photonic researchers to begin studying the thermal behavior of their devices in order to determine their *true* CMOS compatibility. Power dissipation also plays a pivotal role in the tradeoff between optical and copper based interconnects. To be sure, bandwidth of copper interconnects continue improve by using equalization techniques, albeit, at the price of higher power consumption.

Moore's Law is based on the observation that as transistors get smaller, they also become cheaper and faster [17]. They become cheaper because smaller transistors occupy less chip area. Photonic devices must be real estate efficient if they are to be economically compatible with silicon manufacturing. Great progress has been made in reducing the size of silicon photonic devices; however, the progress is running into a fundamental roadblock, namely the diffraction limit of light. The latter states that the minimum waveguide cross section is 200-300nm, and is given by *λ/2n*, where *λ* is the free space wavelength (most commonly 1550nm) and *n* is the refractive index of silicon (~3.5). An attractive solution to the real-estate problem is 3-D integration of electronics and photonics, an example which has recently been demonstrated [19] (Fig. 4). The technology employs a multi-layer SOI structure in which the photonic devices are first formed in the buried silicon layer. The buried layer is patterned by implantation of oxygen through masks residing on the surface layer. The surface silicon layer is then used for conventional CMOS processing.

After years of convergence, silicon and photonics are beginning to diverge! This counter intuitive trend has economic roots. The cost of a mask set for 65nm CMOS process is well over a million USD and is expected to skyrocket in future generations. Costs for design and validation tools are also increasing at alarming rate. The consumer electronics industry can absorb these costs because the large volume of products sold ensures a low production cost per product. With the high volume market for chip scale optical interconnects being at best several years away, the silicon photonics industry is not so lucky. The notion that CMOS manufacturing is synonymous with low cost photonics needs to be revisited.

Fortunately, there are plenty of optical markets that can be served with older CMOS processes that are cost effective. These include 10Gbit/s Ethernet (10GbE) and its long distance equivalent (OC192). The former is a unique opportunity because attempts to extend the reach of copper based one Gbit/s Ethernet to 10GbE have been mostly unsuccessful (unless more expensive cables, than CAT5, are used). In another market, silicon photodetectors are the de-facto standard in x-ray imaging systems used for medical imaging and airport security. Integrating these with electronic amplifiers and signal processing will alleviate the interconnect problems present in high resolution imagers.

There are also applications for silicon photonics beyond data communication. Silicon is already used as mirrors for mid-infrared laser but it also has good nonlinear optical properties at these wavelengths [4]. These plus its large thermal conductivity and high optical damage threshold renders silicon an attractive platform for mid-infrared photonics with applications in medicine, biochemical sensing, and defense. While the research community has judiciously been moving towards ever smaller waveguide dimensions, unique opportunities exist in the opposite extreme of large multi-mode waveguide structures. An example is the recently proposed silicon image amplifier that takes advantage of Raman amplification in the presence of the Talbot effect [20]. Such a device, which is being developed by a UCLA/Northrop Grumman team, is designed to improve the sensitivity of laser based remote imaging systems. These are but a few examples - many more previously unseen applications exist.

An example is optically assisted analog-to-digital conversion (ADC), a technology aimed at solving he bandwidth limitation of electronic ADCs. Among different approaches, the photonic time stretch technique has been the most successful. In this approach a photonic pre-processor slows down the electrical waveform prior to digitization by an electronic ADC [21]. It has recently demonstrated real-time operation at 10 Tsample/s in transient mode [22], and 150 Gsample/s in continuous mode [23]. Silicon photonics provides all the necessary building blocks for producing such as a system on a single chip.

Finally, research into nonlinear optical interactions is leading to unexpected scientific discoveries. As a case in point, we have recently reported that fluctuations of Raman amplified pulses, in the presence of a noisy pump, follow extreme-value statistics - highly non-Gaussian distributions that have been surprisingly successful in describing the frequency of occurrence of extreme events ranging from stock market crashes and natural disasters, the structure of biological systems and fractals, and optical rogue waves [24]. These distributions predict that events much larger than the mean can occur with significant probability, in stark contrast to the ubiquitous Gaussian distribution which heavily favors events close to the mean.

Experimentally, we have found that 16% of the amplified pulses account for 84% of the pump energy transfer, an uncanny resemblance to the empirical 80/20 rule that describes important observation in socioeconomics. A simple mathematical model provides insight into this fascinating behavior and how it emerges in such diverse and seemingly unrelated fields of physical and social sciences.

In summary, silicon photonic is making stunning progress. In terms of commercial success, it stands to have a bright future, as long as the research community recognizes the real challenges that remain and maintains an open mind with respect to its applications.

**Acknowledgments**: The authors work has been supported by the CSWDM and EPIC programs of DARPA-MTO.

**References:**

[1] W. Wayt Gibbs, "Computing at the speed of light," Scientific American, pp. 81-87, November 2004.

[2] B. Jalali, "Making silicon lase," Scientific American, pp. 58-65, February 2007.

[3] R.A. Soref, "Past, present, and future of silicon photonics," Journal of Selected Topics in Quantum Electronics, Vol. 12, pp. 1678-87, November/December 2006.

[4] B. Jalali and S. Fathpour, "Silicon Photonics," Journal of Lightwave

Technology, Vol. 24 (no.12), pp. 4600-15, December 2006.

[5] R. A. Soref, J. Schmidtchen, and K. Petermann, "Large single-mode rib waveguides in GeSi and Si-on-SiO2," Journal of Quantum Electronics, Vol. 27, pp. 1971-74, August 1991.

[6] R.A. Soref and J. Lorenzo, "All-silicon active and passive guided-wave components for λ=1.3 and 1.6 μm," Journal of Quantum Electronics, Vol. 22, pp. 873-9, June 1986.

[7] Z. Huang, N. Kong, X. Guo, M. Liu, N. Duan, A.L. Beck, S.K. Banerjee, and J.C. Campbell, "21-GHz bandwidth germanium-on-silicon photodiode using thin SiGe buffer layers," Journal of Selected Topics in Quantum Electronics, Vol. 12, pp. 1450-54, November/December 2006.

[8] M.A. Foster, A.C. Turner, J.E. Sharping, B.S. Schmidt, M. Lipson, and A.L. Gaeta, "Broad-band optical parametric gain on a silicon photonic chip," Nature, Vol. 441, pp. 960-3, June 2006.

[9] K.K. Tsia, S. Fathpour, B. Jalali, "Electrical control of parametric processes in silicon waveguides," Optics Express, Vol. 16 (no.13), pp. 9838-9843, June 2008.

[10] Y. Liu, C.W. Chow, W.Y. Cheung, and H.K. Tsang, "In-line channel power monitor based on helium ion implantation in silicon-on-insulator waveguides," Photonics Technology Letters, Vol. 18, pp. 1882-84, September 2006.

[11] Y. Liu, et al., IEEE J. Lightwave Technology 2006.

[12] E-K. Tien, N.S. Yuksek, F. Qian, and O. Boyraz, "Pulse compression and modelocking by using TPA in silicon waveguides," Optics Express, pending.

[13] F. Iacona, A. Irrera, G. Franzo, D. Pacifici, I. Crupi, M. Miritello, C.D. Presti, and F. Priiolo, "Silicon-based light-emitting devices: Properties and applications of crystalline, amorphous and Er-doped nanoclusters," Journal of Selected Topics in Quantum Electronics, Vol. 12, pp. 1596-1606, November 2006.

[14] G.Y. Sung, N-M. Park, J-H. Shin, K-H. Kim, T-Y. Kim, K.S. Cho, and C. Huh, "Physics and device structure of highly efficient silicon quantum dots based silicon nitride light-emitting diodes," Journal of Selected Topics in Quantum Electronics, Vol. 12, pp. 1545-55, November 2006.

[15] S. Saito, D. Hisamoto, H. Shimizu, H. Hamamura, R. Tsuchiya, Y. Matsui, T. Mine, T. Arai, N. Sugii, K. Torii, S. Kimura, and T. Onai, "Silicon light-emitting transistor for on-chip optical interconnection," Applied Physics Letters, Vol. 89 (163504), October 2006.

[16] A.W. Fang, H. Park, O. Cohen, R. Jones, M.J. Paniccia, and J.E. Bowers, "Electrically pumped hybrid AlGaInAs-silicon evanescent laser," Optics Express, Vol. 14 (no.20), pp. 9203-10.

[17] D.A. Muller, "A sound barrier for silicon?" Nature Materials, Vol. 4, pp. 645-47, September 2004.

[18] F. Pollack, "New microprocessor challenges in the coming generation of CMOS," ://www.llnl.gov/radiant.

[19] Indukuri, T., Koonath, P. & Jalali, B. Monolithic vertical integration of metal-oxide-

semiconductor transistor with subterranean photonics in silicon. Optical Fiber Comm. Conf., OFC 2006, Anaheim, CA, March 2006.

[20] B. Jalali, V. Raghunathan, and R. Rice, "Multi-mode mid-IR silicon Raman amplifiers," Materials Research Society Symposium, Paper #0958-L11-01, Boston, MA, November 2006.

[21] Y. Han and B. Jalali, "Photonic time-stretched analog-to-digital converter: Fundamental concepts and practical considerations," Journal of Lightwave Technology, Vol. 21 (no.12), pp. 3085-3103, December 2003.

[22] J. Chou, O. Boyraz, D. Solli, and B. Jalali, "Femtosecond real-time single-shot digitizer," Applied Physics Letters, Vol. 91 (no. 16), pp. 161105-07, October 2007

[23] J. Chou, J. Conway, G. Sefler, G. Valley, and B. Jalali, "150 GS/s real-time oscilloscope using a photonic front end," Int'l Topical Meeting on Microwave Photonics (MWP 2008), September 2008.

[24] David Borlaug and Bahram Jalali, "Extreme-value statistics in silicon photonics," Invited Paper, LEOS '08, Newport Beach, CA, November 2008.

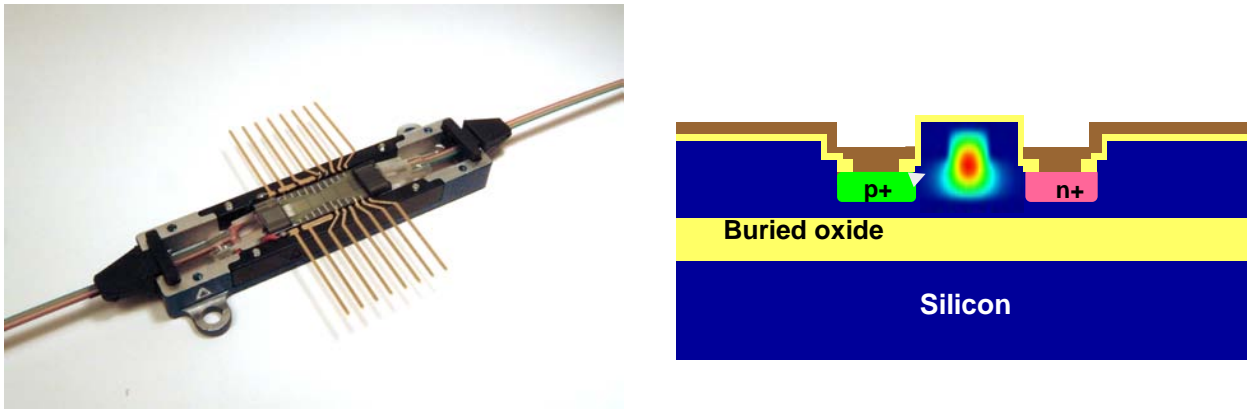

**Figure 2** Left: An 8-channel electrically-variable optical attenuator from Kotura Corporation. The device is used for channel equalization in multi wavelength fiber networks. Right: the active component is a p-n junction straddling a rib-waveguide. Free carrier absorption inside the waveguide core is adjusted by varying the diode current.

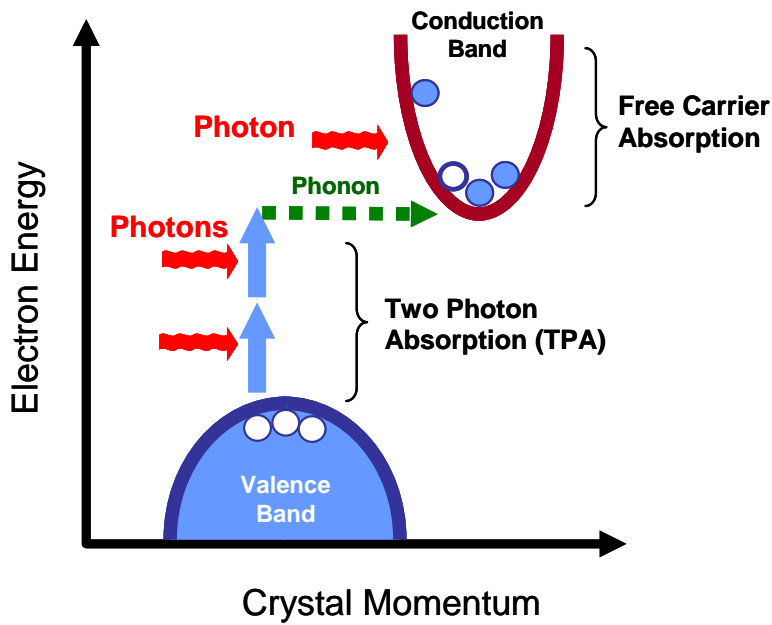

**Figure 1** Two photon absorption (TPA) generates free carriers that in turn cause significant amount of absorption. The loss of photons is main problem in a new class of silicon devices that perform optical amplification, lasing, and wavelength conversion. These devices take advantage of the material's nonlinear optical properties. The high optical intensities that are needed to induce nonlinear effects render silicon lossy.

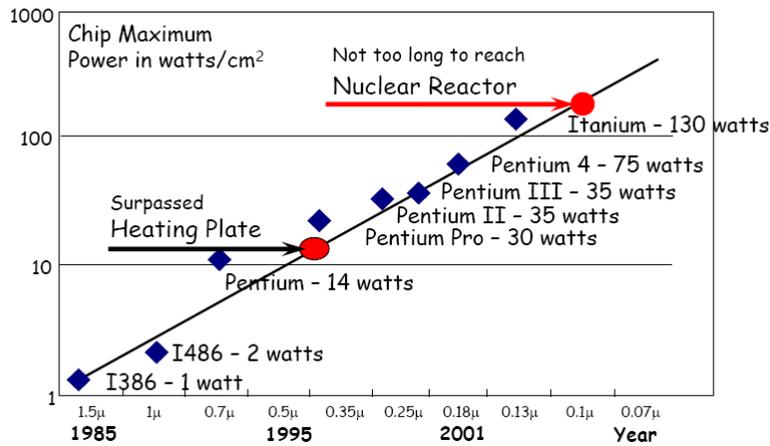

**Figure 3. An unintended consequence of Moore's Law, which implies that transistors will continue to become smaller, cheaper, and faster. Because of rapid increase in transistor count, the power density has already exceeded 100 Watts/cm2. For photonic devices to find their way onto a silicon VLSI chip, they must be low power, and more imprtantly, be able to operate reliably at elevated temperatures as high as 90C.**

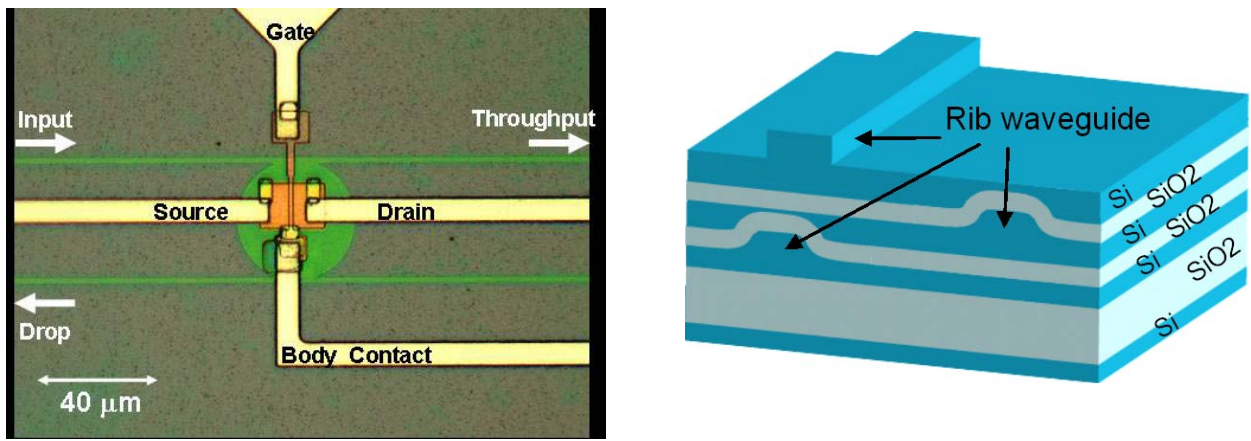

**Figure 4. Right: The ultimate solution to efficient use of wafer real-estate is 3-D integration of electronics on top of buried photonics. The example shown here is an MOS transistor fabricated on top of a sub-surface micro-disk resonator [19]. The fully monolithic process employs multi-layers of SOI, formed by patterned oxygen implantation, in which photonic devices are first formed in the buried silicon layer. Left: 3-D sculpting of microphotonic structures consisting of high index Si and low index SiO2 sections. The process makes use of oxygen implantation through semitransparent masks and has been used to realize vertically coupled photonic devices as well as 3-D integration of MOS transistors on top of buried photonic circuits [19].**